\documentclass[aps.pre,showpacs]{revtex4}
\usepackage{graphicx}
\usepackage{color}
\usepackage{ulem}
\def\be{\begin{equation}}
\def\ee{\end{equation}}     
\def\bfi{\begin{figure}}
\def\efi{\end{figure}}
\def\bea{\begin{eqnarray}}
\def\eea{\end{eqnarray}}

\begin{document}

\title{Fractal character of the phase ordering kinetics of a diluted ferromagnet}

\author{Federico Corberi$^{1,2}$, Leticia F. Cugliandolo$^3$,
Ferdinando Insalata$^{4}$, Marco Picco$^{3}$}
 
\affiliation{$^1$Dipartimento di Fisica ``E.~R. Caianiello'', Universit\`a  di Salerno, 
via Giovanni Paolo II 132, 84084 Fisciano (SA), Italy.
\\
$^2$INFN, 
Gruppo Collegato di Salerno, and CNISM, Unit\`a di Salerno, Universit\`a  di Salerno, 
via Giovanni Paolo II 132, 84084 Fisciano (SA), Italy.
\\
$^3$Sorbonne Universit\'e and CNRS,  
Laboratoire de Physique
Th\'eorique et Hautes Energies, UMR 7589\\
4, Place Jussieu, Tour 13, 5\`eme \'etage,
75252 Paris Cedex 05, France.
\\
$^4$Department of Mathematics, Imperial College London, London SW7 2AZ, United Kingdom.
}

\begin{abstract}

  We study numerically the coarsening kinetics of a two-dimensional
  ferromagnetic system with aleatory bond dilution.
  We show that interfaces between
  domains of opposite magnetisation are fractal on every lengthscale,
  but with different properties at short or long distances. Specifically,
  on lengthscales larger than the typical domains' size the topology
  is that of critical random percolation, similarly to what observed in
  clean systems or models with different kinds of quenched disorder. On smaller
  lengthscales a dilution dependent fractal dimension emerges.
  The Hausdorff dimension increases with increasing dilution
  $d$ up to the value
  $4/3$ expected at the bond percolation threshold
  $d=1/2$. We discuss how such different geometries develop on different
  lengthscales during the phase-ordering process and how their
  simultaneous presence determines the scaling properties of observable
  quantities.

\end{abstract}

\maketitle

\section{Introduction} \label{intro}

The growth of order in a system quenched across a phase transition is a paradigm of slow
relaxation~\cite{Bray94,Puri-Book,Corberi2010}.
A classical example, that we will consider in this paper, is provided by the
coarsening kinetics of a ferromagnet after a quench from an equilibrium state above
the critical temperature to a temperature $T_f$ below it. Considering a uniaxial magnet, for simplicity,
the non-equilibrium evolution is characterised by an early development of domains --
small scale realisation of the two symmetry related low temperature equilibrium phases --
and the subsequent growth of their characteristic linear size $R(t)$.

In the clean cases, that is in the absence
of any source of quenched disorder, such as impurities, external disturbances or lattice defects,
the geometrical properties of ordered regions are well understood~\cite{Bray94,Corberi2008}:
domains are compact objects surrounded  by interfaces which are subjected to thermal roughening.
However, in the small $T_f$ limit and at long times, roughening effects are negligible when observed on the
characteristic scale $R(t)$, and domain walls appear as regular, non fractal objects. 
These features
are reflected by the behaviour of correlation functions, such as the structure factor~\cite{Bray94}.

The geometry of the
growing structure on scales much larger than the typical size of the domains,
instead, is non trivial and has been investigated in two dimensions
recently~\cite{Arenzon07,Sicilia07,Sicilia08,Sicilia09,Blanchard12,Blanchard14,Tartaglia15,Blanchard17,Tartaglia16,Corberi17,Insalata18,Tartaglia18}. 
It was shown that, next to $R(t)$, another length $R_p(t)>R(t)$ is growing faster and,
on scales $R(t)\ll r \ll R_p(t)$ the geometry is that of random percolation at criticality.
This growing percolative structure invades the whole system at the size-dependent time $t_p(L)$
when $R_p(t_p)$ becomes of the order of the linear size $L$ of the sample.
The awareness of percolative features is not only important on its own, since its appearance 
in a strongly correlated system is unexpected and highly non-trivial, but represents also a cornerstone
of one of the few existing analytical theories of
phase-ordering. Indeed, since the distribution of cluster areas is exactly known in critical percolation~\cite{Cardy03},
simply inferring their evolution after the quench allows one to arrive at an exact expression for
the area distribution at any time~\cite{Arenzon07,Sicilia07}. Moreover, the recognition that the systems quickly approach a critical percolation situation after a quench allowed one to understand the results on the percentage of blocked states found after quenches to  zero temperature~\cite{Barros09,Olejarz12,Blanchard13}.

When quenched disorder is present in the system, as it is very often the
case when dealing with real materials, the understanding is
poorer than in the clean cases. It is known from the theoretical study
of model systems that even a tiny amount of quenched noise strongly
inhibits the growth process~\cite{Oh86,Oguz90,BrayHumayun,Puri91,Puri93,Rao93,Rao93b,Oguz94,Gyure95,Fisher98,Fisher01,Corberi02,Paul04,Paul05,Henkel06,Paul07,Aron08,Henkel08,Lippiello10,Park10,Corberi11,Corberi12,Corberi15}. 
This fact was
confirmed in some experimental work on real
materials~\cite{Ikeda90,Schins93,Shenoy99,Likodimos00,Likodimos01}.
The slowing down is accompanied by modifications of the scaling properties of the correlation
functions with respect to those of clean systems~\cite{Corberi15,Likodimos00,Likodimos01}.
Despite all the above, 
it was shown
in~\cite{Corberi17,Insalata18} that basic properties, such as the
compact geometry of the growing domains and their boundaries and the
emergence of the percolative structure on scales beyond $R(t)$,
remain unchanged when going from a clean two-dimensional Ising system to one
with quenched disorder in the form of random fields
or random bonds.

Notwithstanding the interest of the results discussed insofar, a
full characterisation of the geometrical properties of coarsening
in disordered ferromagnets remains a subject to be further pursued.
Among the many open questions, the generality of what
found in~\cite{Corberi17,Insalata18} for the random field Ising model (RFIM) and
the random bond Ising model (RBIM) needs to
be investigated. In particular, the behaviour
of these systems in the very asymptotic regime is unexplored
because the numerical study in~\cite{Corberi17,Insalata18} 
focused on the long-lasting transient regime~\cite{Corberi11,Corberi12} before
the system crosses over to the late non-equilibrium stage at a certain time $t_{cross}$.

In this paper we give a contribution to answering these and related questions
by studying numerically the geometry of coarsening in a system with a different kind of
quenched disorder, the
bond-diluted Ising model (BDIM), namely a ferromagnetic spin system where
pairwise interactions are randomly nullified
with probability $d$, the dilution. We consider this model
because on the one hand $t_{cross}$ is sufficiently short to allow one
to enter the asymptotic stage and, on the other hand, because
it was shown to possess a richer dynamical structure
as compared to different disordered ferromagnetic systems~\cite{Corberi13}.

Our results show profound differences between the 
BDIM and what was found in~\cite{Corberi17,Insalata18} for the
RFIM and the RBIM. These differences regard both the geometric features
of the correlated region on scales $r< R(t)$ and the way in which
the percolative structure develops for $r>R(t)$.

Studying the properties of the average squared winding angle,
a quantity that probes the geometry of the interfaces, we show that,
at variance with the clean case, the domain walls are fractal
with a fractal dimension $D$. This Hausdorff dimension increases with $d$ up to the value
$D=4/3$ expected at the bond percolation threshold $d_c=1/2$.
To the best of our knowledge this is the first time that the
fractality of interfaces has been quantified for this kind of system.
Let us remark that growing domain boundaries are not fractal in the clean IM, the RFIM nor the RBIM for $t<t_{cross}$
(although they may be fractal for $t>t_{cross}$ in cases with quenched randomness). 

Regarding the behaviour at large distances $r>R(t)$ our results
confirm that a spanning cluster with the topology of critical random percolation
also develops in the present model, as in the clean case and in the RFIM and RBIM.
However, what changes here is the dynamics whereby this object develops
and grows. Indeed, while in~\cite{Corberi17,Insalata18} it was
found that $R_p(t)$ is related to $R(t)$ in a way which is independent of
the presence of disorder, on its nature (i.e. whether considering random fields or random bonds)
and on its amount, here such relation depends on $d$. In particular,
at $d=d_c$, where the bond network is itself a critical percolation cluster,
we find that the property $R_p(t)>R(t)$ is no longer obeyed, 
since $R_p(t)$ and $R(t)$ coincide, as it could be expected.

These results, and in particular the dependencies of $D$ and $R_p(t)$
on the amount of disorder $d$, help us clarifying the {\it superuniversal}
properties of coarsening kinetics. Superuniversality~\cite{Fisher88}
can be expressed as the fact that the only effect of quenched disorder
is to slow down the coarsening process, namely the increase of $R(t)$, leaving other properties,
such as the geometric ones, unchanged.
Some numerical simulations gave support to this property in a number of systems~\cite{Paul04,Paul05,Puri91,Puri92,Biswal96,Loureiro10,Gyure95},
 including the RFIM and RBIM for $t<t_{cross}$~\cite{BrayHumayun,Rao93b,Puri93,Sicilia08,Aron08}, 
while other numerical studies pointed against its validity~\cite{Lippiello10,Fisher01,Fisher98,Corberi02,Corberi11,Corberi12}.
In the RFIM and the RBIM the
squared winding angle discussed above does not depend on the presence of
quenched disorder nor on its strength provided that distances are
measured in units of $R(t)$.
However, something different happens in the BDIM, since the effects of disorder cannot
be simply accounted for by $R(t)$, and the scaling functions of observable quantities and
the fractal dimension $D$ of the interfaces are also modified.

This paper is organised as follows: in Sec.~\ref{themodel}
we define the BDIM model and some quantities that will be considered
for its characterisation. In Sec.~\ref{numres} we present and discuss the results of
our simulations. We conclude the Article and discuss some open problems and
future perspectives in Sec.~\ref{concl}.

\section{Model and observable quantities} \label{themodel}

We consider a system with Hamiltonian
\be
{\cal H}(\{S_i\})=-\sum _{\langle ij\rangle}J_{ij}S_iS_j+\sum _i H_i S_i
\; ,
\label{isham}
\ee
where $S_i=\pm 1$ are Ising spin variables located at the vertices of a two-dimensional
$L\times L$ square lattice that we label $i$, with $i=1, \dots, N=L^2$.
$\langle ij\rangle$ denotes nearest 
neighbour sites $i$ and $j$ on the lattice.
In the following we will study the bond-diluted Ising model (BDIM)
where $H_i\equiv 0$ and the $J_{ij}$'s are uncorrelated random variables with
probability $P(J_{ij})=(1-d)\,\delta _{J_{ij},J}+d\,\delta _{J_{ij},0}$, where $\delta$
is the Kronecker function, $J>0$, and $d$ is the dilution. The same Hamiltonian~(\ref{isham})
also describes two different disordered model, the RFIM and the RBIM.
Specifically, for the RFIM one has $J_{ij}\equiv J$ and $H_i=\pm h$ is
uncorrelated in space and sampled from a symmetric bimodal distribution, whereas
for the RBIM, the fields vanish $H_i\equiv 0$ and the ferromagnetic coupling constants
are $J_{ij}=J+\delta _{ij}$,
where $\delta _{ij}$ are independent random numbers extracted from a flat distribution in 
$[-\delta,+\delta]$, with $\delta <J$. The geometrical properties of these two models
have been studied in a previous
paper~\cite{Corberi17} and, in the present analysis, we will often compare their behaviour
to the one of the BDIM, where richer results are found.

Before studying the dynamics of the BDIM, let us review some of its equilibrium properties.
Given the structure of the coupling constants, it is clear that for $d>d_c=1-p_c$, where $p_c$ is
the bond percolation threshold ($p_c=1/2$ in the two-dimensional case considered here)
the bond network is disconnected (in the thermodynamic limit). Hence the spin system is
fragmented into separated parts and a globally ordered (magnetised) state cannot exist at any
temperature, not even at $T=0$. For this reason we will restrict our attention to $d\le d_c$
in the following. In this range of dilution the system undergoes a phase transition at
a critical temperature $T_c(d)$ which decreases from $T_c(0)$ -- the transition temperature
of the usual model without dilution -- to $T_c(d_c)=0$~\cite{Selke98,Heuer92,Kim94,Stinchcombe83}. The coarsening dynamics of a ferromagnetic Ising model with dilution was studied in~\cite{Burioni06,Burioni07,Burioni13,Corberi15,Park10,Paul07,Iwai93}.

A dynamic update is introduced  by reversing spins with the Glauber transition rates
\be
w(S_i\to -S_i)=\frac{1}{2}\left [1-S_i\tanh \left (\frac{H^W_i}{T}\right )\right ]
\label{trate}
\ee
where $H^W_i$ is the local Weiss field
\be
H^W_i=\sum _{j\in  nn(i)}J_{ij}S_j
\; ,
\ee
and the sum runs over the nearest neighbours 
$nn(i)$ of $i$. Here and in the following we measure temperature in
units of the Boltzmann constant.

We will consider an instantaneous quench in which 
the system is prepared at time $t=0$ in an infinite temperature equilibrium state,
namely an uncorrelated one with zero magnetisation, and it is subsequently evolved by attempting
spin flips according to the transition rates (\ref{trate}) evaluated
at the final temperature $T=T_f$ of the quench. Periodic boundary conditions will be adopted. 

Let us now define the various observables that we will consider in our study.
The average size of the growing domains $R(t,d)$ will be computed as~\cite{Bray94,Puri-Book,Corberi2010}
\be
R(t,d)=L^{2}[E(t,d)-E_{eq}(d)]^{-1}
\; ,
\label{defrt}
\ee
where $E(t,d)=\langle H(t)\rangle$ is the non-equilibrium average (namely taken over
different initial condition and thermal histories) of the energy at time $t$, and $E_{eq}(d)$ is
the energy of the equilibrium state at $T_f$. Equation~(\ref{defrt}) is a standard method
to determine the typical size of the domains in clean systems~\cite{Bray94} and is well
suited to be applied also to diluted systems, as discussed in~\cite{Corberi13,Corberi15}.

The wrapping probability $\pi (t,d)$ is the probability that at time $t$ a connected wrapping
cluster of equally aligned spins is present in the system. 
The wrapping cluster can cross the system
in different ways. It can span the system from one side
to the other horizontally or vertically. The probabilities of these events
will be denoted as $\pi_{1,0}(t,d)$ and $\pi_{0,1}(t,d)$, respectively.
On the square lattice one has $\pi_{0,1}(t,d)=\pi_{1,0}(t,d)$.
A cluster can also span the sample in {\it both}  horizontal
and vertical directions,  with probability $\pi_{0,0}(t,d)$.
Finally, clusters can percolate along one of the two diagonal directions with
equal probabilities $\pi_{1,-1}(t,d)$ and $\pi_{1,1}(t,d)$.
With periodic boundary conditions the torus can be wrapped more than once
but this occurs with extremely small probability and we will not consider these
spanning configurations.

The fractal properties of the domain boundaries can be studied by considering
the (average) squared winding angle
$\langle \theta ^2(r,t,d)\rangle$ defined as follows: at a given time, for any two points $i,j$ 
on the external perimeter 
of a cluster we compute the angle $\theta _{ij} $ 
(measured counterclockwise) between the tangent to the perimeter in $i$ and the one in $j$.
Upon repeating the procedure for all the couples of perimeter points at distance $r$, 
taking the square and averaging over the non-equilibrium ensemble, 
$\langle \theta ^2(r,t,d) \rangle$ is obtained. For numerical convenience we have considered
only the largest cluster in the system.

Another quantity we will analyse is the pair connectedness function $C(r,t,d)$.
In percolation theory~\cite{Stauffer,Saberi,Christensen} this quantity is defined as
the probability that two points
at distance $r$ belong to the same cluster. In spin systems
we use the same definition where, in this case, two spins belong to
the same cluster if they are
aligned and are connected by a path of aligned spins. 
In order to compute this quantity at a given time we
first identify all the
domains of positive and negative spins. Then we evaluate $C(r,t)$ as
\be
C(r,t,d)=\frac{1}{4 L^2}\sum _i \sum _{i_r} \langle \delta _{S_i,S_{i_r}} \rangle
\; ,
\label{connect} 
\ee
where $\delta _{S_i,S_j}=1$ if the two spins belong to the same cluster and $\delta _{S_i,S_j}=0$
otherwise, and $i_r$ is a site at distance $r$ from $i$. 

\section{Numerical results} \label{numres}

\begin{figure}[h]
\vspace{1.9cm}
\centering
\rotatebox{0}{\resizebox{.95\textwidth}{!}{\includegraphics{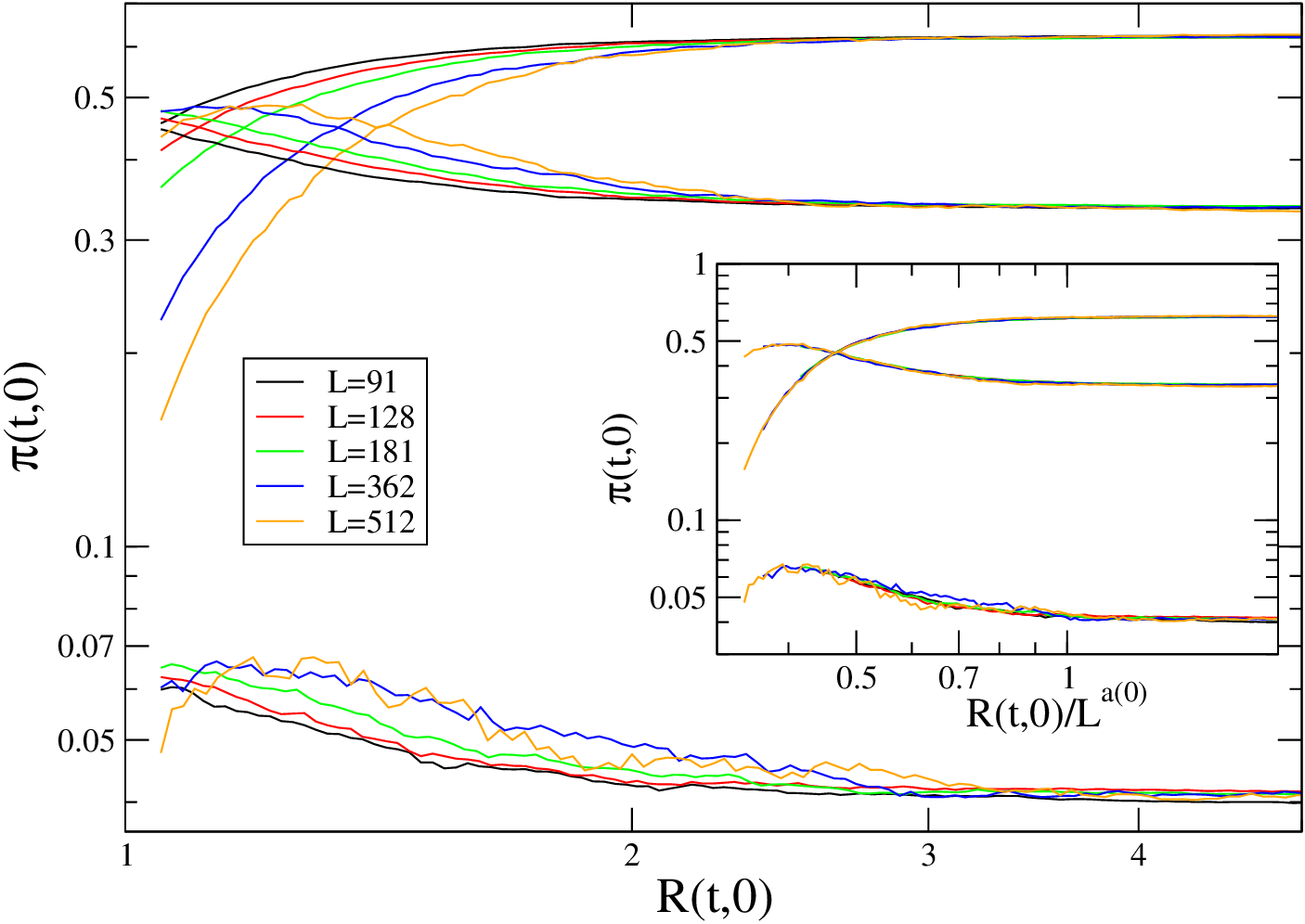}}}
\caption{The wrapping probabilities $\pi(t,0)$ are plotted against $R(t,0)$ with double logarithmic
scales, for $d=0$, the clean model. The curves reaching the largest saturation value correspond to the
probability $\pi _{0,0}(t,0)$ (different curves are for different system sizes, see the colour code in the key).
The curves reaching the lowest asymptotic value represent the probability $\pi_{1,-1}(t,0)+\pi_{1,1}(t,0)$.
Finally, the remaining curves that approach an intermediate value close to 0.338 
corresponds to $\pi_{0,1}(t,0)+\pi_{1,0}(t,0)$. In the inset the same quantities are plotted
against $R(t,0)/L^{a(0)}$, with $a(0)=0.178$.}  
\label{fig_probcross_pure}
\end{figure}

\begin{figure}[h!]
\vspace{2cm}
\centering
\rotatebox{0}{\resizebox{.95\textwidth}{!}{\includegraphics{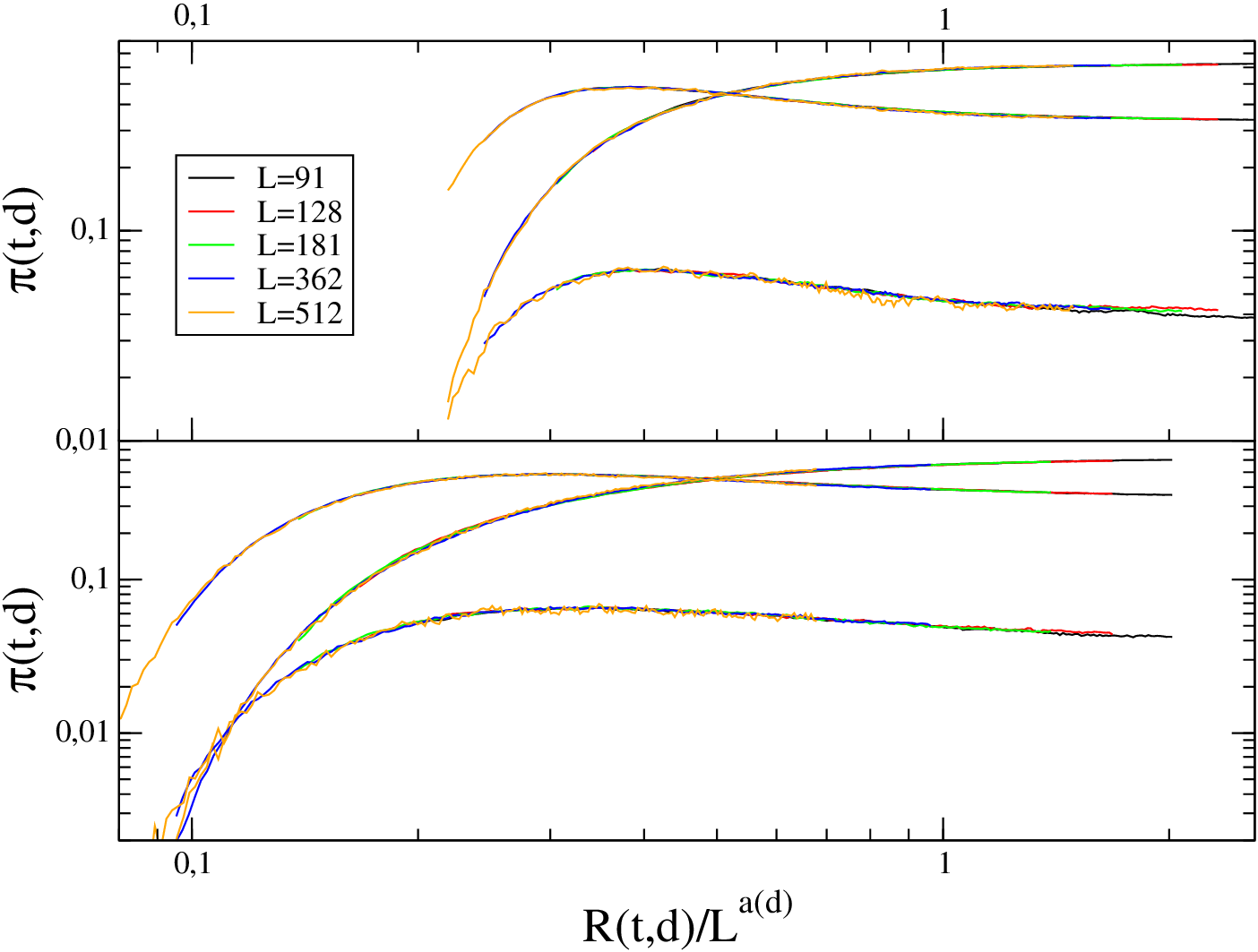}}}
\caption{The wrapping probabilities $\pi(t,d)$ are plotted against $R(t,d)/L^{a(d)}$ with double logarithmic
    scales, for $d=0.25$ (upper panel, $a(0.25)=0.32$) and $d=0.4$
    (lower panel, $a(0.4)=0.54$).
    The  curves reaching the largest saturation value show the
probability $\pi _{0,0}(t,d)$ (different curves correspond to different system sizes, see the 
colour code in the key).
The curves reaching the lowest asymptotic value display the probability $\pi_{1,-1}(t,d)+\pi_{1,1}(t,d)$
and the other ones correspond to $\pi_{0,1}(t,d)+\pi_{1,0}(t,d)$.}  
\label{fig_probcross_diluted}
\end{figure}

In this Section  we present and discuss the outcome of our numerical simulations.
These have been performed on systems with different sizes, from $91\times 91$ up to
$512\times 512$, with periodic boundary
conditions, and an average over $10^4-10^5$ realisations 
has been taken. The data presented in the following are usually obtained for the largest
size ($512\times 512$), while smaller sizes are used to control finite size effects,
see for instance Figs.~\ref{fig_probcross_pure} and \ref{fig_probcross_diluted} and the related discussion.
We have performed quenches with $J=2$ to $T_f=0.75$ and $T_f=1$ and we found similar
results in the two cases. Both these temperatures are below $T_c$ in the whole range
of dilutions considered in our simulations.
In the following, we will discuss data for $T_f=0.75$ unless otherwise stated.

\begin{figure}[h]
\vspace{2cm}
\centering
\rotatebox{0}{\resizebox{.95\textwidth}{!}{\includegraphics{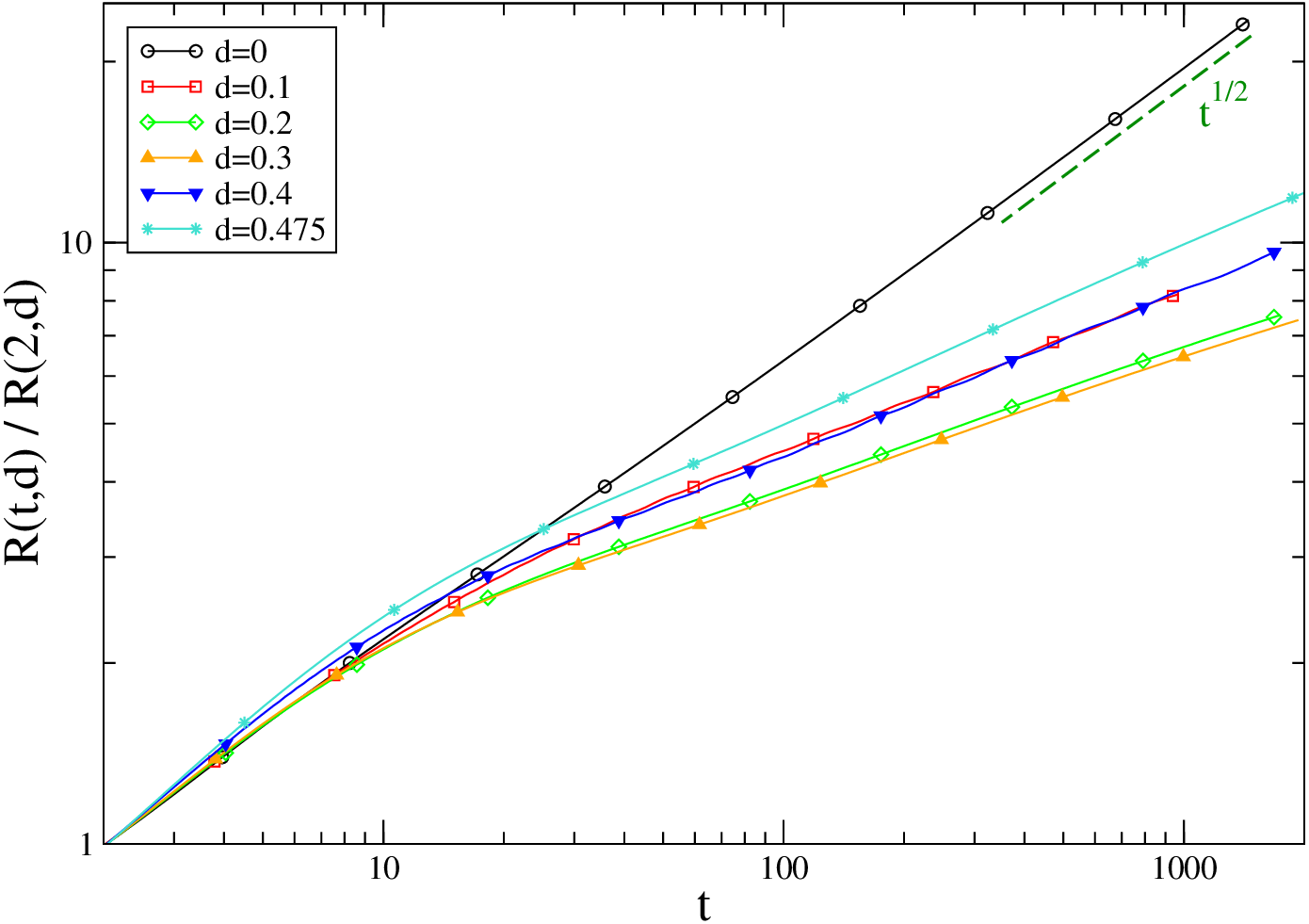}}}
\caption{$R(t,d)/R(2,d)$ against time $t$ after the quench, for different choices of $d$, see the key. Dividing by $R(2,d)$ allows to
better compare curves with different $d$. The dashed green line is the behaviour $t^{1/2}$ expected for 
the pure system with $d=0$. 
}  
\label{fig_R}
\end{figure}

In Fig.~\ref{fig_R} we show  the characteristic length $R(t,d)$ after the quench
of systems with various dilutions. The behaviour of this quantity has been thoroughly discussed
in~\cite{Corberi13,Corberi15} and in~\cite{Corberi15d} for two and three dimensional systems, respectively, 
with both site or bond dilution. The main feature we want to stress here is the strong
dependence of the growth law on the amount of disorder, a feature that is usually
observed in ferromagnetic systems with any kind of quenched disorder. In particular, the usual law 
$R(t,0)\sim t^{1/2}$ is observed for the clean case only. The growth slows down upon increasing $d$ up 
to values of order $d=0.2-0.3$ and then it gets progressively faster as $d$ is further raised.
This non-monotonic behaviour is interpreted in~\cite{Corberi13,Corberi15,Corberi15d} as due to the
relatively fast growth at $d=d_c$ induced by the percolative structure of the bond
network. Notice, however, that the kinetics is slower than in the clean case for any $d>0$.  

Let us now discuss the behaviour of the wrapping probabilities. 
These quantities are exactly known
for two-dimensional critical percolation under periodic boundary conditions.
They are~\cite{Pinson} 
\begin{eqnarray}
\pi_{0,1}+\pi_{1,0}&\simeq& 0.3388 \; , \nonumber \\
\pi_{0,0}&\simeq& 0.61908  \; , 
\label{spanpperc}
 \\
\pi_{1,-1}+\pi_{1,1} &\simeq& 0.04196 \; . \nonumber
\end{eqnarray}
The behaviour of the wrapping probabilities during the phase-ordering
process in the case without dilution is plotted in Fig.~\ref{fig_probcross_pure}.
In this figure, and in the following ones, we use $R(t,d)$ as a natural reparametrization
of time. 
The wrapping probabilities start from zero immediately after the quench,
because the initial state is a random mixture of an equal amount of up and down spins and
there cannot be any spanning cluster in such a configuration given that the site percolation
threshold ($\simeq 0.59...$) is larger than the fraction 1/2 of both kinds of spins.
At long times (large $R(t,d)$) the crossing probabilities saturate
to values which are very precisely consistent with the ones  
at critical percolation given in Eq.~(\ref{spanpperc}), as
pointed out in \cite{Barros09,Olejarz12}. This fact is a first confirmation that
the spanning object is a percolation cluster.
These asymptotic values are attained around a certain value of $R$ that can
be interpreted as $R(t_p,d)$, where $t_p$ is the time when the growing percolative structure
hits the system boundaries, namely $R_p(t_p,d)=L$. 

Comparing curves relative to systems with different size 
one observes that $R(t_p,d)$ increases with $L$.
This is because, as already discussed, the cluster which spans the system
at $t_p$ does not appear altogether at that particular time but its size $R_p$ gradually grows  
until it crosses the entire system.
One finds that a collapse of the wrapping probabilities for different system sizes is 
achieved by plotting the $\pi$'s against $R(t,d)/L^{a(d)}$, and the best collapse
is obtained for $a(0)=0.178$ in this case, as shown in the inset of Fig.~\ref{fig_probcross_pure}.

Before moving on to the study of the wrapping probabilities in the presence of dilution it is useful to 
summarise what is known for other kinds of quenched disorder, namely in the case
of the RFIM and the RBIM. It was shown in~\cite{Corberi17} that the presence of 
disorder does not change the qualitative nor the quantitative behaviour of the $\pi$'s. 
Indeed, by fixing the system size $L$ and plotting the crossing probabilities against $R(t)$, it was found that the curves
corresponding to the clean case, the RFIM and the RBIM, fall one on top of the other, 
despite the fact that,
as in the dilute case, the addition of frozen randomness greatly slows down the kinetics.
The superposition is observed for any
strength of the quenched disorder, namely for different values of $h$ and $\delta$. This property,
sometimes referred to as {\it superuniversality}~\cite{Fisher88}, 
implies that the sole effect of disorder is to change the form of $R(t)$, leaving other properties unmodified. 

On the other hand, by varying the system size $L$, one can again obtain data collapse for the wrapping
probabilities by plotting them against $R(t)/L^a$, where $a $ is an exponent which turns
out to be independent both on the kind and strength of disorder.

Let us now  consider the system at hand in this paper.  
When dilution is present, the qualitative features of the wrapping probability do not
change, as it can be seen in Fig.~\ref{fig_probcross_diluted}. This is similar to what was observed in the 
RFIM and the RBIM. In particular, the limiting values of the $\pi$'s are the ones given in Eq.~(\ref{spanpperc}),
a fact which again confirms the percolative nature of the spanning clusters.
However, rescaling the curves with the value of the exponent $a(0)=0.178$ of the clean case
does not produce any superposition of the curves for different system sizes. A good collapse can be
still obtained, but using a $d$-dependent exponent $a(d)$, as it can be seen in Fig.~\ref{fig_probcross_diluted}
in the cases with $d=0.25$ and $d=0.4$ (data collapses of comparable quality are obtained for any $d$, with
appropriate values of $a(d)$).

\begin{figure}[h!]
\vspace{2cm}
\centering
\rotatebox{0}{\resizebox{.5\textwidth}{!}{\includegraphics{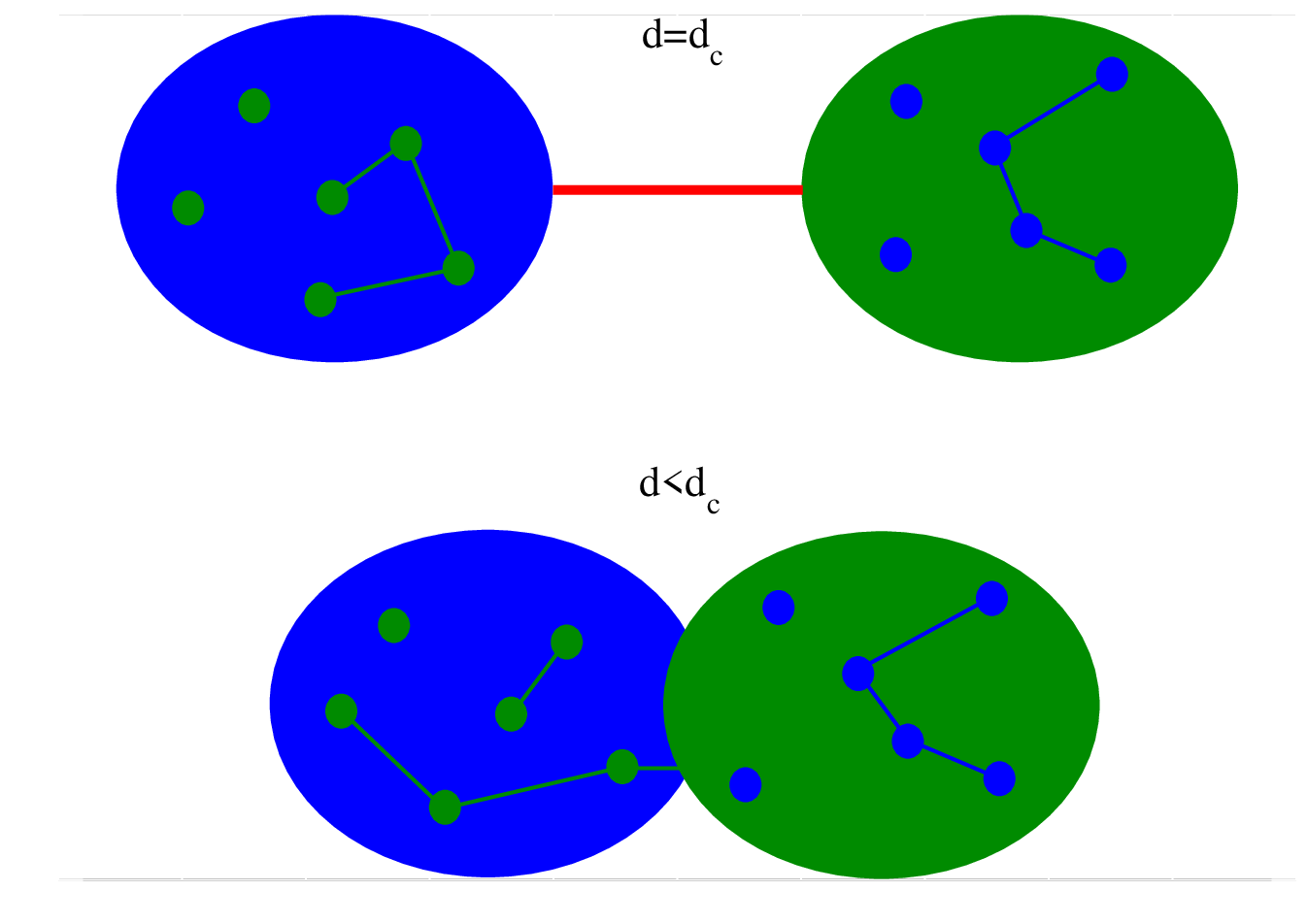}}}
\caption{Pictorial representation of the phase-ordering process at $d=d_c$ (upper part) or $d<d_c$ (lower part). The blue and green regions 
are domains with spins up and down. The red line connecting them for $d=d_c$ is a red bond. See the text for a thorough explanation of the meaning of this sketch. }  
\label{fig_redbonds}
\end{figure}

The fact that the exponent $a(d)$ must depend on $d$ is suggested by
an argument according to which $a(d_c)=1$, which is very different from $a(0)=0.178$. The argument is the following.
Right at $d=d_c$ the network of the $J_{ij}$'s is a percolation structure. It is well known that such a fractal is characterised by
the presence of so-called {\it red bonds}, namely bonds whose removal results in the disconnection of the network. 
This geometry is pictorially sketched in the upper part of Fig.~\ref{fig_redbonds}. 
It is also known~\cite{Burioni06,Burioni07,Burioni13,Corberi15} that, during phase-ordering kinetics at low temperatures on such a structure, interfaces between domains of opposite spins are located, 
with large probability, on the red bonds, where they get pinned for a very long time $\tau _{pin}$. Indeed, after depinning they
quickly travel in a time $\tau_{move}\ll \tau_{pin}$ towards the next red bond. Since $\tau_{pin}/\tau _{move}\to \infty$ for
$T_f\to 0$, in the low-temperature limit configurations in which the interface is located 
away from a red bond can be
neglected if one is interested in the typical (average) configurations of the system at a generic time.
In Fig.~\ref{fig_redbonds} the blue region on the left represents
a group of, say, up spins, and the green region is a group of down spins. Inside such regions small islands
of reversed spins can be present, and we draw them schematically by circles that can be connected among them (this is rendered by
lines). 
The two large regions (blue and green) are linked by a red bond where the interface is located. Clearly
i) the two blobs correspond to domains of aligned spins (possibly with thermal fluctuations in their interiors)
and hence their size is of order $R$ and ii) the size ${\cal R}$ of the percolation cluster
cannot be larger than the size $R$ of the domains because, in order to do that, the (say) left region should be connected
through the red bond to some down spin inside the right region, but in this case the interface would not be located
on the red bond.
This implies that $R$ and ${\cal R}$ coincide and hence $a(d_c)=1$.
Notice that the situation is
very different for $d<d_c$, where red bonds do not exist and the two domains are now connected by a number of order $R$ of links
instead that by a single red bond, as represented
in the lower part of the figure. In this case the percolation 
cluster can extend beyond $R$. For instance the down spins of the right domain could be connected to some of the islands of reversed
spins inside the up domain, as shown in the figure. In this configuration the size $R$ of the correlated domains remains
of the order of the typical size of the two blobs, but ${\cal R}$ can extend much beyond $R$. 

\begin{figure}[h!]
\vspace{2cm}
\centering
\rotatebox{0}{\resizebox{.95\textwidth}{!}{\includegraphics{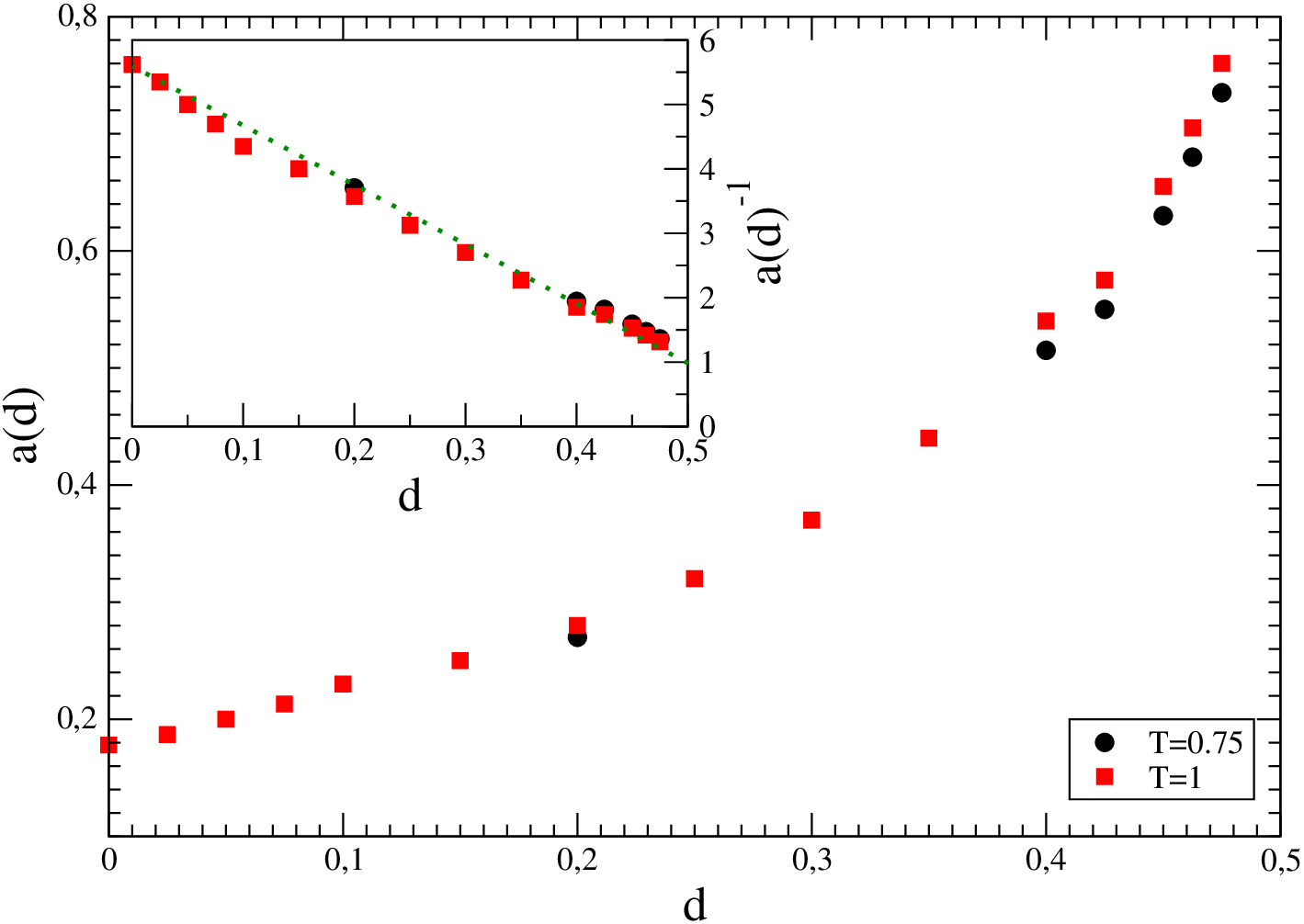}}}
\caption{The exponent $a(d)$ against $d$, for $T_f=0.75$
  and $T_f=1$ (see the key). In the inset we plot $a(d)^{-1}$ against $d$. The dotted green line 
  represents $a(d)^{-1}=a(0)^{-1}-cd$, with $c=2(a(0)^{-1}-1)$.}  
\label{fig_gamma2}
\end{figure}

The dependence of $a$ on $d$ is shown in Fig.~\ref{fig_gamma2} for two temperatures.
From the comparison between these two cases one concludes that $a(d)$ is rather independent
of $T_f$. These results imply that the formation of a spanning percolative cluster occurs
at progressively larger values of $R$ as $d$ is increased. In other words, dilution
slows down this process.
Moreover, data show that the result of the previous argument, namely $a(1)=1$,
is consistent with the numerical outcome. Let us mention that in Ref.~\cite{Blanchard14} it was conjectured
that, on deterministic lattices,
$a \propto n_c^{-1}$, where $n_c$ is the average coordination  number of the network.
Since in our diluted case $n_c(d)=n_c(0)(1-d)$, a blind extension of the conjecture above 
to the random lattice at hand would result in the linear behaviour $a(d)^{-1} \propto n_c(0)(1-d)$.
Although this form does not strictly describe the data, nevertheless, a linear decrease
of $a(d)^{-1}$ with $d$ is neatly observed. Indeed, in the inset of
Fig.~\ref{fig_gamma2} one sees that a behaviour 
of the form $a(d)^{-1}=a(0)^{-1}-cd$ (dotted greed line), where $c=2(a(0)^{-1}-1)$
is a constant, is well consistent with the numerical results. 

\begin{figure}[h!]
\vspace{2cm}
\centering
\rotatebox{0}{\resizebox{.95\textwidth}{!}{\includegraphics{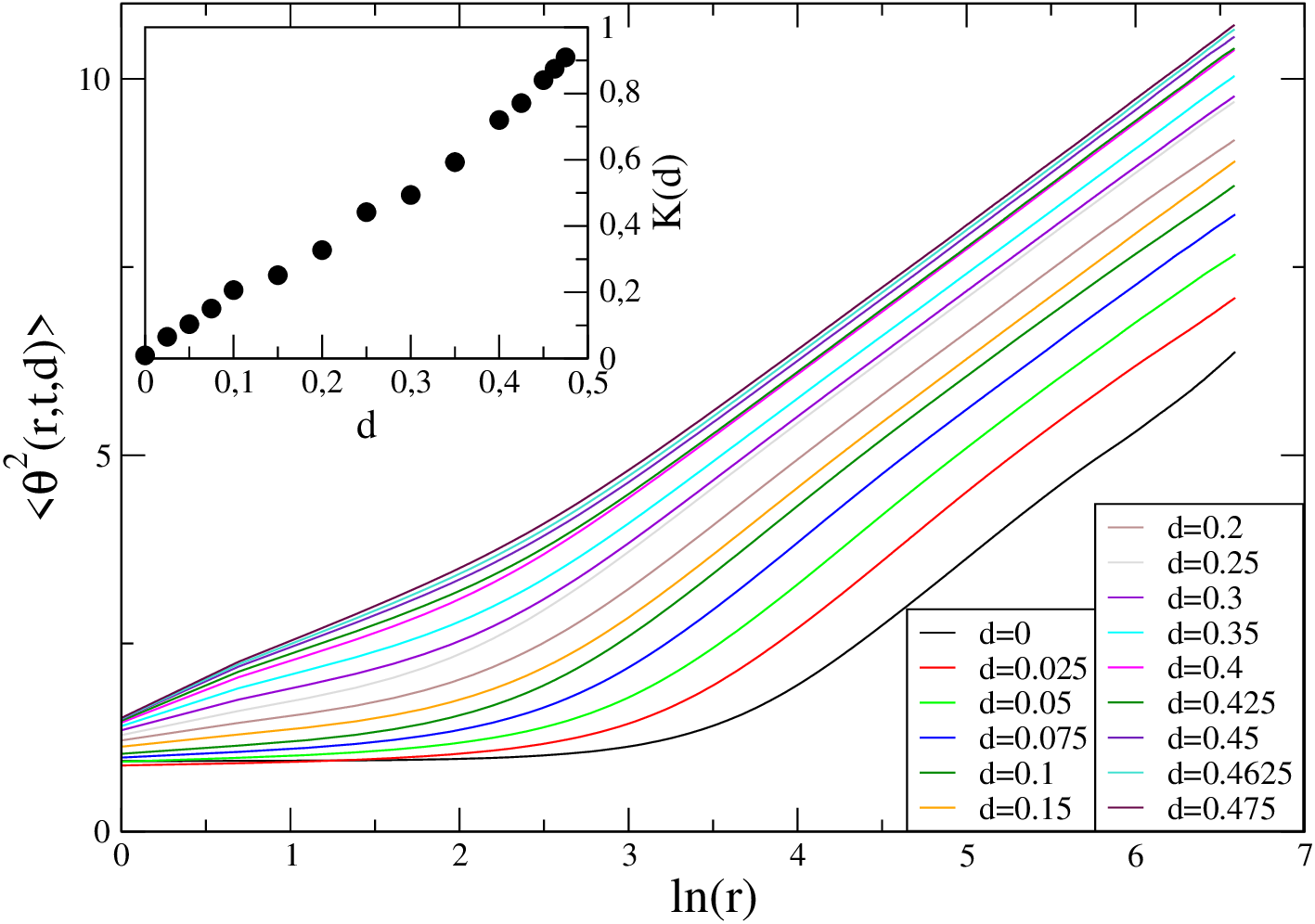}}}
\caption{The averaged squared winding angle $\langle \theta ^2(r,t,d)\rangle$ is plotted against $\ln r$ for different
  values of the dilution $d$ (see the key). For each $d$ the data refer to the longest
  simulated time (this time is $t=1218$, 1218, 939, 1125, 939, 1680, 1681, 939, 1865,
  2142, 1681, 2603, 2603, 2603, 2603 for $d=0, 0.25,\dots,0.475$, respectively). In the
  inset the slope $K(d)$ of the small $r$ part of the curves is plotted against $d$.
  This quantity was computed by a linear fitting of the curves in the range $0.6\le \ln r \le 1.4$.}
\label{fig_winding_varid}
\end{figure}

The next quantity we consider is the averaged squared winding angle of the spin cluster interfaces.
Its behaviour is exactly known in $2d$ critical percolation 
at $p=p_c$~\cite{Duplantier,Wilson}, where one has
\be
\langle \theta ^2_{perc}(r,r_0) \rangle =a+K \ln \left ( \frac{r}{r_0}\right )
\, .
\label{windexact}
\ee
$a$ is a non-universal constant and $r_0$ is the lattice spacing.
Beyond percolation theory, Eq.~(\ref{windexact}) is valid for the winding angle 
of a self similar interface, the  fractal dimension of which, $D$, 
is related to $K$ by $D=4/(4-K)$.
$\langle \theta ^2\rangle$ was measured on the interfaces of the phase-ordering domains 
in the clean Ising model in~\cite{Corberi17}. It was found that the form~(\ref{windexact}) is obeyed,
with $R(t,0)$ playing the role of $r_0$, and 
with two different values of $K$ on scales $r\ll R(t,0)$ and $r\gg R(t,0)$. For $r\ll R(t,0)$ one probes the
geometrical properties of the interfaces of the correlated domains. Since these are compact objects with a 
regular interface one has $D=1$, and hence $K=0$~\cite{nota}. On the other hand, for $r\gg R(t,0)$ the fractal nature
of the interface of the percolating structure emerges, leading to $K=K_{perc}$. This can be seen in 
Fig.~\ref{fig_winding_varid} by looking at the curve with $d=0$. In the case of the RFIM and the RBIM,
once the data for the winding angle are plotted against $r/R$, one finds
a perfect superposition of the curves for the disordered models with those of the clean case~\cite{Corberi17}.
This is similar to what was observed for the crossing probabilities and is, again, due to the superuniversality property.

Let us now consider the case with dilution. The main difference with the behaviour of the IM, RFIM and RBIM
is that in the regime $r\ll R(t,d)$ one 
finds $K=K(d)$, with $K(d)$ an increasing function of $d$ (see the inset of Fig.~\ref{fig_winding_varid}). This means that 
the interfaces of the correlated domains which are growing are fractal. 

The following argument suggests that for $r \ll R(t,d)$, $K(d_c)=1$.
Indeed, up to now, we have considered the so-called spin or geometrical clusters built by connecting nearest neighbouring 
sites occupied by the same spin value.
For the bond percolation corresponding to the dilution $d=d_c$, the natural objects to consider are the Fortuin-Kasteleyn (FK) 
clusters~\cite{FK} for which the $K$ of their interface is $K=K^{\rm FK}_{int}=12/7$ and, therefore, the fractal dimension
is $D^{\rm FK}_{int}=4/(4-K^{\rm FK}_{int}) = 7/4$. 
However, we are studying here the interfaces of the spin clusters~\cite{Delfino} and we are therefore measuring a $K$ (and its 
associated fractal dimension) that will not take this value.  
In fact, we are measuring a $K$ that has been shown to be equal to the one of the {\it 
external perimeter} of the FK clusters~\cite{Duplantier2,Adams}. 
For any critical $Q$ state Potts model, and critical percolation is a particular case with $Q=1$, the fractal dimensions of the 
FK clusters interface, $D^{\rm FK}_{int}$,  and the one of the FK clusters external perimeter, $D^{\rm FK}_{extp}$, are related by the equation 
$(D^{\rm FK}_{int}-1) ( D^{\rm FK}_{extp}-1) = 1/4$~\cite{Duplantier2}. 
Replacing $D^{\rm FK}_{int}= 7/4$ one readily finds
$D^{\rm FK}_{extp} = 4/ 3$ and $K^{\rm FK}_{extp} = K(d_c) = 1$.  
We see in the inset of Fig.~\ref{fig_winding_varid} that, indeed, data are
consistent with $K(d_c)=1$ which then implies that the fractal dimension of the interfaces is $D=4/3$ at $d=d_c$. 
Notice also that  $K(d)$ increases in an approximately linear way with $d$, a fact for which we have no explanation.

In the large distance regime, for $r\gg R(t,d)$, one recovers the slope $K=K_{perc}=12/7$, that is independent of 
$d$ and the same as the one observed for the clean system, which shows quite convincingly that
the geometry of the clusters of aligned spins on such large scales is the one of percolation.

\begin{figure}[h!]
\vspace{2cm}
\centering
\rotatebox{0}{\resizebox{.95\textwidth}{!}{\includegraphics{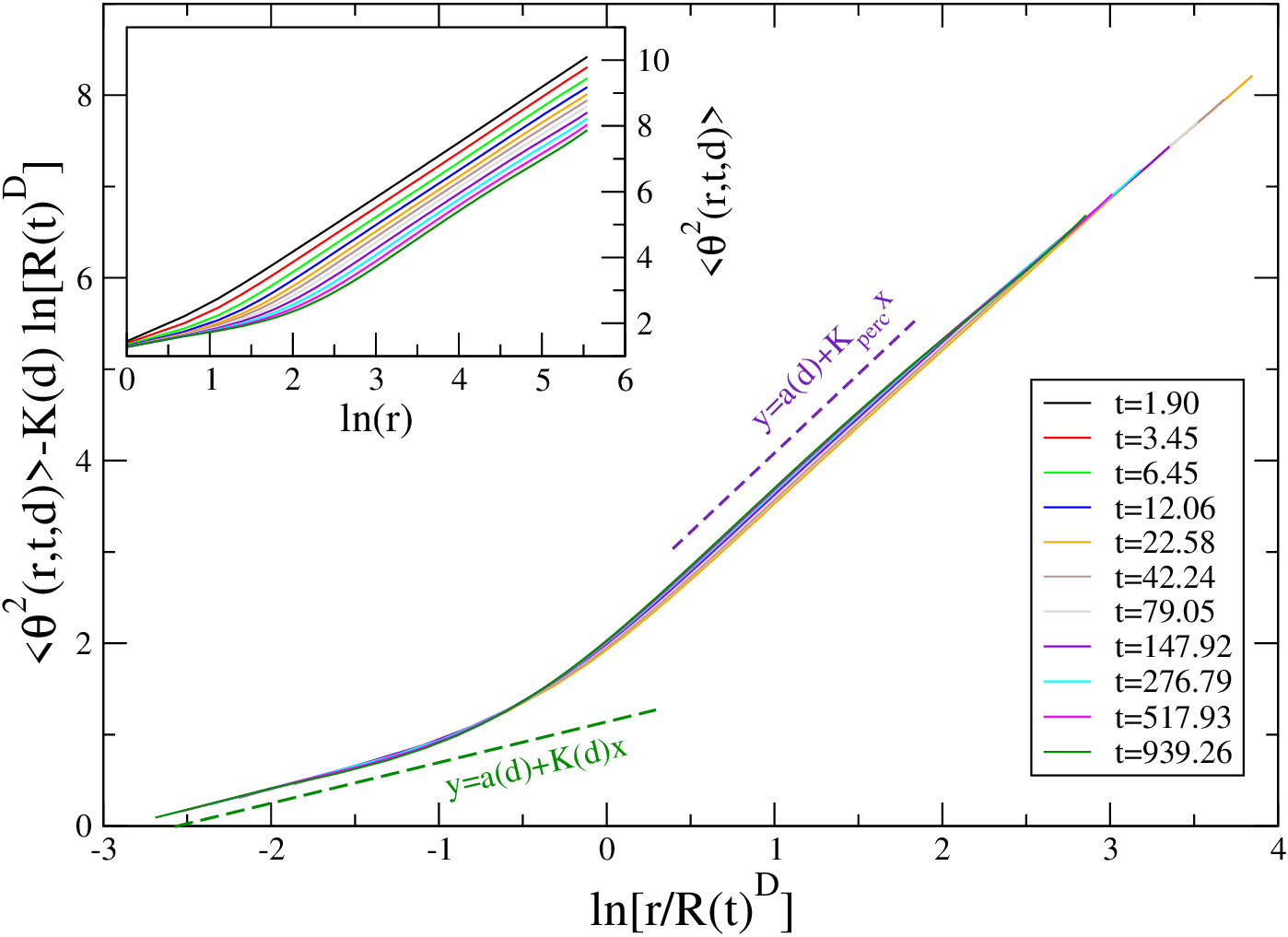}}}
\caption{In the inset $\langle \theta ^2(r,t,d)\rangle$ evaluated at different times $t$ (see the key)
  is plotted against $\ln r$ for a system with $d=0.25$. In the main part of the figure,
  for the same sets of data of the inset (but only for $t\ge 22.58$, in order to focus
  on the asymptotic behaviour), we plot
  the quantity $\langle \theta ^2(r,t,d)\rangle-K(d)\ln [R(t,d)^D]$
  (see Eq.~(\ref{scal_winding})) against $\ln[r/R(t,d)^D]$. The dashed green and violet
  lines are the linear behaviours with slopes $K(d)$ and $K_{perc}$, respectively.
  }  
\label{fig_winding_d025_resc}
\end{figure}

The behaviour of $\langle \theta ^2\rangle$ for a fixed value of $d$ and different times is
shown in the inset of Fig.~\ref{fig_winding_d025_resc} (a similar behaviour is obtained
for all values of $d$ below $d_c$). Here we see that the value
$r_{cross}$ of $r$
where there is a change in the slope of the curve increases as time elapses. Since
$R(t,d)^D$ is a natural length associated to the fractal character of the interfaces
we assume, as in Ref.~\cite{Tartaglia18}, that $r_{cross}\sim R(t,d)^D$. Hence, recalling the discussion above, we argue that $\langle \theta ^2 \rangle$ obeys the following form
\be
\langle \theta ^2(r,t,d)\rangle=\left \{
\begin{array}{ll}
  a(d)+K(d)\ln r  & \qquad \mbox{for} \quad r\ll R(t,d)^D \; ,  \\
  a(d)+K(d)\ln [R(t,d)^D]+K_{perc} \ln \left [\frac{r}{R(t,d)^D}\right ]  & \qquad \mbox{for} \quad r \gg R(t,d)^D \; .
\end{array} \right .
\label{wa_expr}
\ee
This formula simply expresses the fact that the squared winding angle
has two linear behaviours
with slopes $K(d)$ and $K_{perc}$ which match at $r=R(t,d)^D$. Clearly, for $r\simeq R(t,d)^D$,
$\langle \theta ^2\rangle$ interpolates smoothly between the two limiting forms
of Eq.~(\ref{wa_expr}). According to this conjecture one should have
\be
\langle \theta ^2(r,t,d)\rangle-K(d)\ln [R(t,d)^D]=f_d\left [\frac{r}{R(t,d)^D}\right ]
\label{scal_winding}
\ee
with
\be
f_d(x)=\left \{
\begin{array}{ll}
  K(d) \; \ln x  & \qquad \mbox{for} \quad x\ll1 \; , \\
  K_{perc} \, \ln x  & \qquad \mbox{for} \quad x\gg 1 \; . 
\end{array} \right .
\label{scal_f_winding}
\ee
Equation~(\ref{scal_winding}) implies that plotting 
$\langle \theta ^2(r,t,d)\rangle-K(d)\ln [R(t,d)^D]$
as a function of $r/R(t,d)^D$ for each value of $d$ should lead to data collapse of the
curves at different times on the mastercurve with the properties~(\ref{scal_f_winding}).
We check this feature in Fig.~\ref{fig_winding_d025_resc}. Here, in the main part of the
figure, we find a very good collapse of the curves, and the limiting behaviour of the
mastercurve agrees with those in Eq.~(\ref{scal_f_winding}). Similar results are found
for any value of $d$. Notice, however, that the mastercurve $f_d(x)$ depends on $d$.
As a last remark, let us point out that the non trivial fractal character of the interfaces due to dilution, 
which gives rise to $K=K(d)$, is observed up to scales as large as $R^D>R$. This is at variance with possible effects due to 
thermal roughening, see the discussion in~\cite{nota}, which are confined to very 
short length scales. It is, instead, similar to what we observed in two other cases: (i) the quench to a critical point, in particular, the one of the clean Ising model~\cite{Blanchard12} and (ii) the dynamics of the voter model~\cite{Tartaglia15,Tartaglia18}.

Let us now turn  to the properties of the pair connectedness defined in Eq.~(\ref{connect}).
For large distances $r$, random percolation theory at $p=p_c$ in a $d=2$ infinite system
gives~\cite{Stauffer,Saberi,Christensen}
\be
C_{perc}(r,r_0)\sim \left ( \frac{r}{r_0}\right )^{-2\Delta}
\label{connectperc}
\ee
where $r_0$ is a microscopic length, e.g. the lattice spacing, and with
the critical exponent $\Delta =5/48$. Since we are considering a system with periodic boundary conditions which corresponds to a torus, then for a system of finite size $L$, the form~(\ref{connectperc}) contains some correction of order ${r/ L}$ which will cause an upward bending of the curve with respect to the algebraic behaviour~\cite{Corberi17}.

\begin{figure}[h!]
\vspace{2cm}
\centering
\rotatebox{0}{\resizebox{.95\textwidth}{!}{\includegraphics{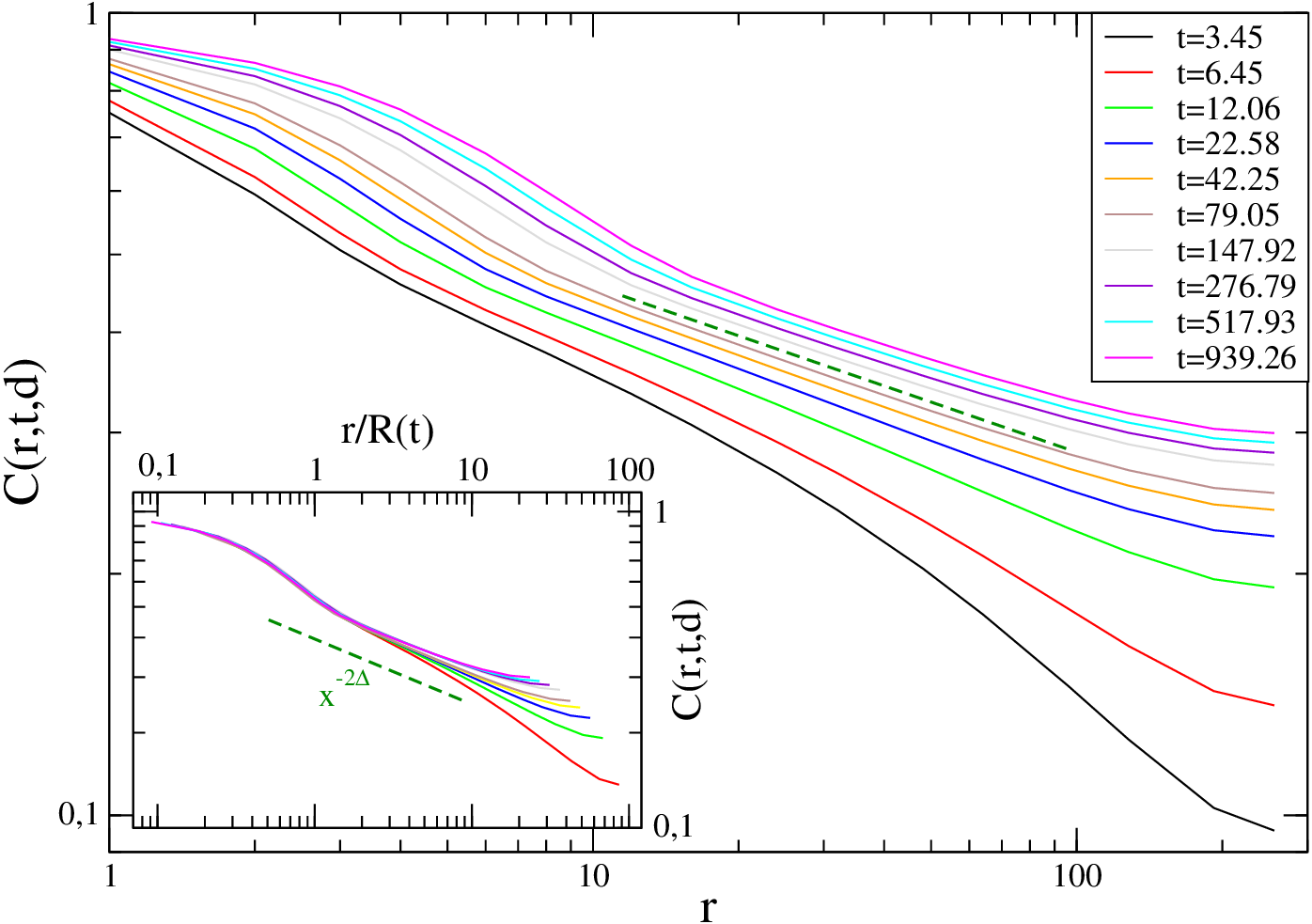}}}
\caption{The pair connectedness function $C(r,t,d)$ is plotted against $r$ at different times
  (see the key) for $d=0.25$. The dashed green line is the algebraic behaviour $x^{-2\Delta}$ 
  of critical percolation
  Eq.~(\ref{connectperc}). In the inset the same data (except the one at the earliest time) are
plotted against $r/R(t)$.}  
\label{fig_connect_d025}
\end{figure}

\begin{figure}[h!]
\vspace{2cm}
\centering
\rotatebox{0}{\resizebox{.95\textwidth}{!}{\includegraphics{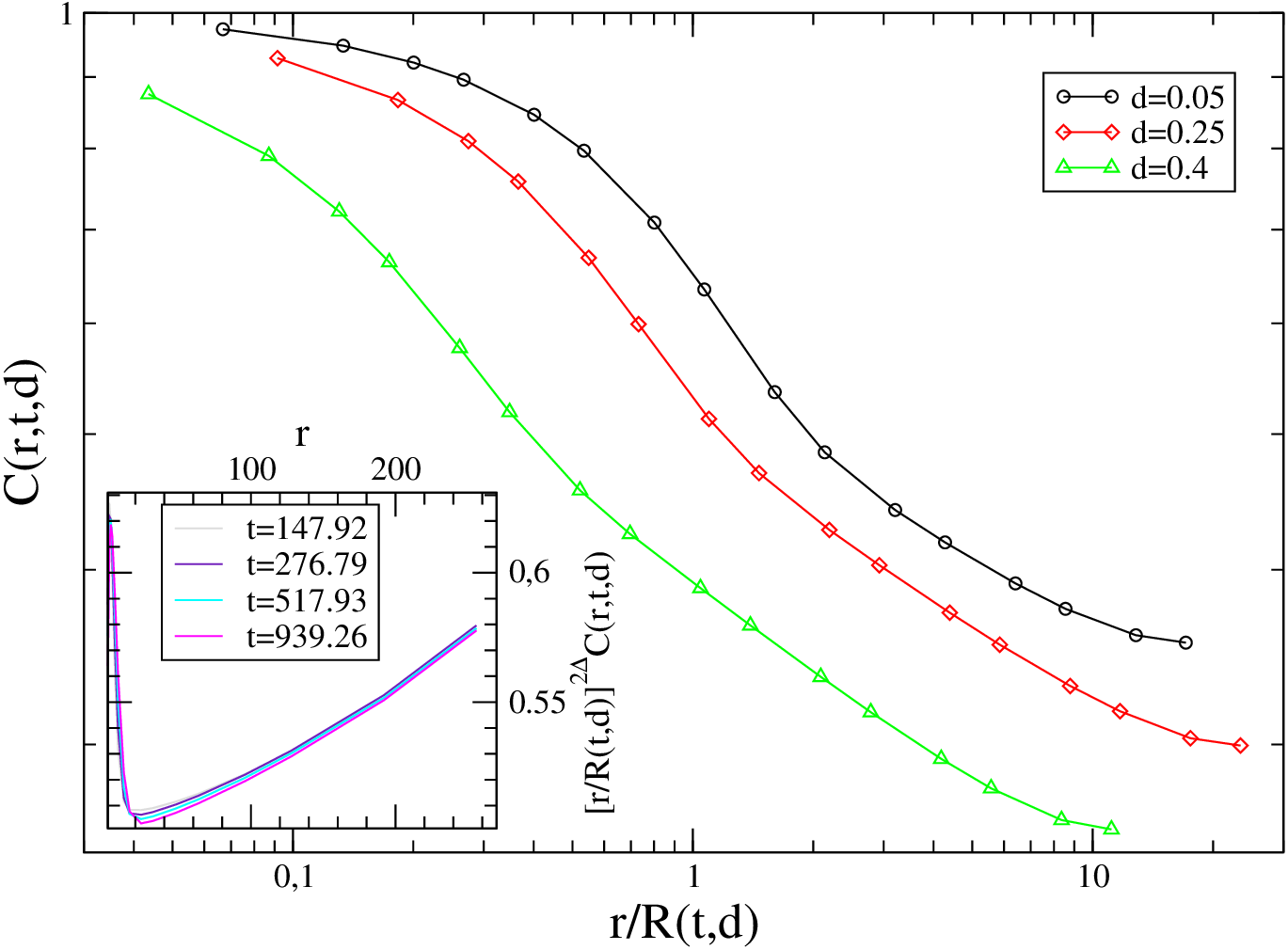}}}
\caption{The pair connectedness function $C(r,t,d)$ at the longest simulated time
  is plotted against $r/R(t,d)$ for $d=0.05$, $d=0.25$, and $d=0.4$ (see the key).
  In the inset, the data for $d=0.25$ are presented by plotting $[r/R(t,d)]^{2\Delta}C(r,t,d)$
against $r$, in order to collapse the finite size part of the curves.
}  
\label{fig_connect_compare}
\end{figure}

In Fig.~\ref{fig_connect_d025} we plot the pair connectedness against $r$ measured in the coarsening
stage of the diluted model with $d=0.25$ at different
times (similar results are found for other values of $d$). Qualitatively, the behaviour of these curves is analogous to the one that was found in the
clean case and in the RFIM and RBIM~\cite{Corberi17}.
A first observation is that, since the number of domains decreases during coarsening,
the probability that two randomly chosen points belong to the same cluster, namely the area
$\sum _rC(r,t,d)$ below the curves, increases in time. Secondly,
the behaviour of the pair connectedness is different for $t\ll t_p$ and $t\gg t_p$.
Indeed, for $t\gg t_p$ the percolating structure has established for $r>R(t,d)$, whereas
it is not well developed for $t\ll t_p$. This is reflected by the fact that the slope
$r^{-2\Delta}$ is clearly observed for sufficiently large values of $r$ in the curves for
$t\gtrsim 42.25$, whereas a faster decay occurs at early times. Let us notice also
that the $r^{-2\Delta}$ behaviour is spoiled at very large $r$ or more precisely for finite values of ${r \over L}$
since our simulations are done on a torus geometry. 
For $t\gg t_p$ the pair connectedness crosses over
between a short distance behaviour, for $r\ll R(t,d)$, where the properties inside the correlation distance
are tested, to the large distance behaviour, $r\gg R(t,d)$, where there is no correlation and the
percolative structure emerges. In this regime one expects Eq.~(\ref{connectperc}) to hold,
with $r_0=R(t,d)$~\cite{Corberi17}.
This behaviour can be summarised in the scaling form
\be
C(r,t,d)={\cal C}_d\left (\frac{r}{R(t,d)}\right ),
\label{scalc}
\ee
valid for an infinite system, with
\be
{\cal C}_d(x)\sim x^{-2\Delta}
\label{scal_f_connect}
\ee
for $x\gg 1$.
In the inset of Fig.~\ref{fig_connect_d025} we plot the same data of the main figure
against $r/R(t,d)$, in order to check the accuracy of
Eqs.~(\ref{scalc}) and (\ref{scal_f_connect}). In this plot one clearly sees that the curves
change behaviour around $r/R(t,d)=1$. For $r/R(t,d)<1$ one has a good collapse
at all times, even for $t<t_p$. This is because the growth of correlations on scales
$r<R(t,d)$ is independent of
the establishment of percolation on scales much larger than $R(t,d)$.
For $r/R(t,d)>1$ one expects data collapse only for $t>t_p$ and in a range of $r$
free from finite size effects. This is indeed observed in the inset of Fig.~\ref{fig_connect_d025}.
Notice that the finite size effects limit the range over which collapse is observed to
smaller and smaller values of $r/R(t,d)$ in this kind of plot. Let us also mention that
one can collapse the curves for $r/R(t,d)>1$ with a different rescaling~\cite{Corberi17}.
Indeed, finite size scaling implies that in a system of size $L$ the form~(\ref{scal_f_connect}) 
changes into ${\cal C}_d(x)\simeq x^{-2\Delta}\Lambda (r/L)$, where $\Lambda (x)$ is a scaling function
describing the finite size properties. Hence, by plotting
$[r/R(t,d)]^{2\Delta} C(r,t,d)$ against $r/L$ one should get data collapse for
$r\gg R(t,d)$ at different times and for different system sizes.
This is shown to be true in the inset of Fig.~\ref{fig_connect_compare}
(here, since the system size $L$ is fixed we plot simply against $r$).

A final important remark is the fact that, as already observed for the squared winding
angle, the scaling function ${\cal C}_d$ does actually depend on $d$. This implies that
superuniversality does not hold for this quantities. We show this in the main part of
Fig.~\ref{fig_connect_compare}, where we plot $C(r,t_{max},d)$, where $t_{max}$ is the
longest simulated time, against $r/R(t,d)$ (this is the best determination of
${\cal C}_d$ apart from finite size effects), for $d=0.05$, $d=0.25$ and $d=0.4$.
This figure shows a marked difference between the scaling functions as $d$ is varied,
despite a certain similarity among the curves.

\section{Conclusions} \label{concl}

In this paper we studied numerically the phase-ordering kinetics of the two-dimensional BDIM with Glauber transition rates,
i.e. with non conserved order parameter dynamics, quenched from an equilibrium initial state at infinite temperature
to a low temperature $T_f$. The growth law and other dynamical properties of this system, 
with similar choices of the model parameters,
had been considered in~\cite{Corberi15}, and in the three dimensional case in~\cite{Corberi15b}. The closely related model with
site dilution was analysed in~\cite{Corberi13}. In all these studies it was argued that the growth law is logarithmically
slow for any $0<d<d_c$. For the limiting values of the dilution, instead, the system behaves differently. 
With $d=0$ there is no disorder and the usual algebraic law $R(t,0)\sim t^{1/2}$ holds. The case $d=d_c$, instead, is
much less trivial, because the fractal character of the bond network (or site network in the case of site dilution), 
results in an algebraic growth of $R(t,d_c)$ regulated by a temperature dependent exponent.

In this paper we continue the study of the coarsening kinetics of diluted systems focusing on the characterisation of the
geometrical properties of the dynamic configurations of the quenched BDIM on the square lattice. Beyond addressing
some of the properties of the domains and their boundaries, an issue that has been longly discussed in 
system with quenched disorder
~\cite{Nattermann87,Fisher86,Kardar87,Halpin89,Huse85,Kardar85,Huse85b,Corberi16,Corberi16b}, we also investigate the geometrical
properties on scales much larger than those of the correlated regions, following a more recent research line aimed
at the comprehension and characterisation of the emerging percolative features.    

We find that domain walls are fractal interfaces with a Hausdorff dimension that depends continuously on the
dilution parameter $d$. This is at variance with what is expected for the
RBIM~\cite{Nattermann87,Fisher86,Kardar87,Halpin89}
where the roughness of the interfaces
due to the bond randomness is expected to be described by a unique exponent, irrespective of the amount of disorder.
This is at odds with the fact the the BDIM is an instance of a system with bond disorder, as the RBIM, and one could have expected the same exponent in the two cases (and independently of the strength of disorder) since both belong to the ``interface disorder" family of systems.
  
Our results confirm the development, on scales larger than $R(t,d)$, of an uncorrelated object with the topology of a critical
percolation cluster. The size $R_p(t,d)$ of this object grows faster (for any $d<d_c$) than $R(t,d)$, in a way which is regulated
by a dilution-dependent exponent. Interestingly, right at $d_c$, $R_p(t,d)\propto R(t,d)$, a fact which is possibly due to the
percolative nature of the bond network itself.   
  
This study allows us to discuss the issue of superuniversality in this system. Our results clearly indicate that, at variance with
what was found for the RFIM and the RBIM considered in Refs.~\cite{Corberi17,Insalata18}, this property is not obeyed 
(at least not fully) in the BDIM.
This conclusion is reached by the recognition that geometrical features such as $D$ and $R_p$ show a continuous dependence
on the amount of dilution. Indeed, our study shows that physical quantities such as, for instance, the crossing probabilities, 
the averaged squared winding angle and the pair connectedness function obey some instance of scaling form, but with
scaling functions that are themselves dependent on the dilution parameter $d$. We expect similar results to hold for
the related model with site dilution.

 \vspace{1cm}
 
 \noindent
 {\bf Acknowledgements}
L. F. C. is a member of Institut Universitaire de France. We warmly thank our former students T. Blanchard and A. Tartaglia for several years of collaborative research on these topics. F.C. acknowledges financial support by MIUR project PRIN2015K7KK8L.


\begin{thebibliography}{0}

\bibitem{Bray94} 
A. J. Bray, Adv. Phys. {\bf 43}, 357 (1994).

\bibitem{Puri-Book}
{\it Kinetics of Phase Transitions}, edited by S. Puri and V. Wadhawan, (CRC Press, Boca Raton, 2009). 

\bibitem{Corberi2010}
   F. Corberi, L. F. Cugliandolo, and H. Yoshino, Growing length scales in aging
systems in: {\it Dynamical heterogeneities in glasses, colloids, and granular media}, edited by L. Berthier, G. Biroli, J.-P. Bouchaud, L. Cipeletti,
and W. van Saarloos (Oxford University Press, Oxford, 2010).

\bibitem{Corberi2008}
  F. Corberi, E. Lippiello, and M. Zannetti,  Phys. Rev. E {\bf 78}, 011109 (2008).

\bibitem{Arenzon07}  
J. J. Arenzon, A. J. Bray, L. F. Cugliandolo, and A. Sicilia, Phys. Rev. Lett. {\bf 98}, 145701 (2007).

\bibitem{Sicilia07}
A. Sicilia, J. J. Arenzon, A. J. Bray, 
and L. F. Cugliandolo, Phys. Rev. E {\bf 76}, 061116 (2007).

\bibitem{Sicilia08}
A. Sicilia, J. J. Arenzon, A. J. Bray, and L. F. Cugliandolo,
EPL {\bf 82}, 10001 (2008).

\bibitem{Sicilia09}
A. Sicilia, Y. Sarrazin, J. J. Arenzon, A. J. Bray, 
and L. F. Cugliandolo, Phys. Rev. E {\bf 80}, 031121 (2009).

\bibitem{Blanchard12} 
T. Blanchard, L. F. Cugliandolo, and M. Picco,
J. Stat. Mech. P05026 (2012).

\bibitem{Blanchard14}  
T. Blanchard, F. Corberi, L. F. Cugliandolo, and M. Picco, EPL {\bf 106}, 66001 (2014).  

  \bibitem{Tartaglia15}
  A. Tartaglia, L. F. Cugliandolo, and M. Picco,  Phys. Rev. E {\bf 92}, 042109 (2015). 

  \bibitem{Tartaglia16}
  A. Tartaglia, L. F. Cugliandolo, and M. Picco,  Europhys. Lett. {\bf 116}, 26001 (2016).  
  
 \bibitem{Blanchard17}
  T. Blanchard, A. Tartaglia, L. F. Cugliandolo, and M. Picco,  J. Stat. Mech. 113201 (2017).  
  
\bibitem{Corberi17}
 F. Corberi, L. F. Cugliandolo, F. Insalata, and M. Picco, Phys. Rev. E {\bf 95},  022101 (2017).

\bibitem{Insalata18}
  F. Insalata, F. Corberi, L. F. Cugliandolo, and M. Picco,  J. Phys.: Conf. Ser. {\bf 956},
  012018 (2018). 
  
  \bibitem{Tartaglia18}
  A. Tartaglia, L. F. Cugliandolo, and M. Picco,  J. Stat. Mech. 083202 (2018).

\bibitem{Cardy03}  
J. Cardy and R. M. Ziff, J. Stat. Phys. {\bf 110}, 1 (2003).  

\bibitem{Barros09} 
K. Barros, P. L. Krapivsky, and S. Redner, Phys. Rev. E {\bf 80}, 040101 (2009).

\bibitem{Olejarz12}  
J. Olejarz, P. L. Krapivsky, and S. Redner, Phys. Rev. Lett. {\bf 109}, 195702 (2012).

\bibitem{Blanchard13}  
T. Blanchard and M. Picco, Phys. Rev. E {\bf 88},  032131 (2013).

\bibitem{Oh86} 
J. H. Oh, and D. Choi, Phys. Rev. B {\bf 33}, 3448
(1986).

\bibitem{Oguz90} 
E. Oguz, A. Chakrabarti, R. Toral, and J. D. Gunton, Phys. Rev. B {\bf 42}, 704 (1990). 

 \bibitem{BrayHumayun}
A. J. Bray and K. Humayun, J. Phys. A {\bf 24} , L1185 (1991).

\bibitem{Puri91}  
S. Puri, D. Chowdhury, and N. Parekh, J. Phys. A {\bf 24}, L1087 (1991). 

\bibitem{Puri93} 
S. Puri and N. Parekh, J. Phys. A {\bf 26}, 2777 (1993). 

\bibitem{Rao93} 
M. Rao and A. Chakrabarti, Phys. Rev. E {\bf 48}, 
R25(R) (1993).
 
\bibitem{Rao93b} 
M. Rao and A. Chakrabarti, Phys. Rev. Lett. {\bf 71}, 3501 (1993). 

\bibitem{Oguz94} 
E. Oguz, J. Phys. A {\bf 27}, 2985 (1994).
 
\bibitem{Gyure95} 
M. F. Gyure, S. T. Harrington, R. Strilka, and H. E. Stanley, Phys. Rev. E {\bf 52}, 4632 (1995).

\bibitem{Fisher98}     
D. S. Fisher, P. Le Doussal, and C. Monthus,
Phys. Rev. Lett. {\bf 80}, 3539 (1998).

\bibitem{Fisher01} 
D. S. Fisher, P. Le Doussal, and C. Monthus, Phys. Rev. E
{\bf 64}, 066107 (2001). 

\bibitem{Corberi02} 
F. Corberi, A. de Candia, E. Lippiello, and M. Zannetti, Phys. Rev. E {\bf 65}, 046114 (2002).

\bibitem{Lippiello10}
  E. Lippiello, A. Mukherjee, S. Puri, and M. Zannetti, Europhys. Lett. {\bf 90},
  46006 (2010).

\bibitem{Paul04} 
R. Paul, S. Puri, and H. Rieger, Europhys. Lett. {\bf 68}, 881 (2004). 

\bibitem{Paul05} 
R. Paul, S. Puri, and H. Rieger,
Phys. Rev. E {\bf 71}, 061109 (2005).

\bibitem{Henkel06} 
M. Henkel and M. Pleimling, Europhys. Lett. {\bf 76}, 561 (2006). 

\bibitem{Paul07}
R. Paul, G. Schehr, and H. Rieger, 
Phys. Rev. E  {\bf 75}, 030104(R) (2007).

\bibitem{Aron08} 
C. Aron, C. Chamon, L. F. Cugliandolo, and M. Picco, J. Stat. Mech. P05016 (2008).

\bibitem{Henkel08} 
M. Henkel and M. Pleimling, Phys. Rev. B {\bf 78}, 224419 (2008). 

\bibitem{Park10} 
H. Park and M. Pleimling, Phys. Rev. B {\bf 82}, 144406 (2010).

\bibitem{Corberi11} 
F. Corberi, E. Lippiello, A. Mukherjee, S. Puri, and M. Zannetti, J. Stat. Mech.  P03016 (2011).

\bibitem{Corberi12} 
F. Corberi, E. Lippiello, A. Mukherjee, S. Puri, and M. Zannetti, Phys. Rev. E {\bf 85}, 021141 (2012).

\bibitem{Corberi15}  
F. Corberi, R. Burioni, E. Lippiello, A. Vezzani, and M. Zannetti, Phys. Rev. E {\bf 91}, 062122 (2015).


\bibitem{Ikeda90} 
H. Ikeda, Y. Endoh, and S. Itoh, Phys. Rev. Lett. {\bf 64}, 1266 (1990).

\bibitem{Schins93}   
A. G. Schins, A. F. M. Arts, and H. W. de Wijn, Phys. Rev. Lett.
{\bf 70}, 2340 (1993).

\bibitem{Shenoy99} 
D. K. Shenoy, J. V. Selinger, K. A. Gr\"uneberg, J. Naciri,
and R. Shashidhar, Phys. Rev. Lett. {\bf 82}, 1716 (1999).

\bibitem{Likodimos00}   
V. Likodimos, M. Labardi, and M. Allegrini, Phys. Rev. B
{\bf 61}, 14440 (2000).

\bibitem{Likodimos01}  
V. Likodimos, M. Labardi, X. K. Orlik, L. Pardi, M. Allegrini, S. Emonin, and O. Marti, Phys. Rev. B {\bf 63}, 064104 (2001).



\bibitem{Corberi13} 
F. Corberi, E. Lippiello, A. Mukherjee, S. Puri, and M. Zannetti, Phys. Rev. E {\bf 88}, 042129 (2013).


\bibitem{Fisher88} 
D. S. Fisher and D. A. Huse, Phys. Rev. B {\bf 38}, 373 (1988).


\bibitem{Stauffer}
D. Stauffer and A. Aharony, 
{\it Introduction to Percolation Theory} (Taylor and Francis, London, 1994).

\bibitem{Christensen}
K. Christensen, {\it Percolation Theory} (Imperial College Press, London, 2002).

\bibitem{Saberi}
A. A. Saberi,
Phys. Rep. {\bf 578}, 1 (2015).

\bibitem{Corberi15c}  
F. Corberi, Comptes rendus - Physique {\bf 16}, 332 (2015). 

\bibitem{Puri92} 
S. Puri and N. Parekh, J. Phys. A {\bf 25}, 4127
(1992). 

\bibitem{Bray91} 
A. J. Bray and K. Humayun, J. Phys. A {\bf 24}, L1185 (1991). 

\bibitem{Biswal96} 
B. Biswal, S. Puri, and D. Chowdhury, Physica A {\bf 229}, 72 (1996).

\bibitem{Castellano98}
C. Castellano, F. Corberi, U. Marini Bettolo Marconi, and A. Petri, Journal de Physique IV  {\bf 8},93 (1998).

\bibitem{Corberi15b} 
F. Corberi, E. Lippiello, and M. Zannetti,
J. Stat. Mech. P10001 (2015). 

\bibitem{Imry75} 
Y. Imry and S.-k. Ma, Phys. Rev. Lett. {\bf 35}, 1399 (1975).

\bibitem{Kolton}
J. L. Iguain, S. Bustingorry, A. B. Kolton, and L. F. Cugliandolo,
Phys. Rev. B  {\bf 80},  094201 (2009).

\bibitem{Pinson}  
H. Pinson,
J. Stat. Phys. {\bf 75}, 1167 (1994).

\bibitem{Duplantier}
B. Duplantier and H. Saleur,
Phys. Rev. Lett. {\bf 60}, 2343 (1988).

\bibitem{Wilson} 
B. Wieland and D. B. Wilson, 
Phys. Rev. E {\bf 68},  056101  (2003).


 \bibitem{Corberi15d} F. Corberi, E. Lippiello and M. Zannetti,  J. Stat. Mech. P10001 (2015).  
 
 \bibitem{Burioni06}
 R. Burioni, D. Cassi, F. Corberi, and A. Vezzani, Phys. Rev. Lett. {\bf 96}, 235701 (2006).
 
 \bibitem{Burioni07}
 R. Burioni, D. Cassi, F. Corberi, and A. Vezzani, Phys. Rev. E {\bf 75}, 011113 (2007).

 \bibitem{Burioni13}
 R. Burioni, F. Corberi, and A. Vezzani, Phys. Rev. E {\bf 87}, 032160 (2013).

  
  \bibitem{nota}
It should be mentioned that, in principle, if the quench is made to a finite temperature $T_f$, roughening of interfaces
takes place and, on scales of the order of the roughening length ${\cal R}$, this should result in $K\ne 0$. 
However, since in $d=2$ the roughness of the interface is of order
${\cal R}(t)\simeq a(T_f)R(t)^{1/2}$~\cite{Corberi2008,Corberi16}, where $a(T_f)$ is a constant which vanishes for $T_f \to 0$,
this effect would produce $K\ne 0$ on scales ${\cal R}(t)\ll R(t)$ so small that the effect cannot be observed in our data
(see Fig.~\ref{fig_winding_varid}).         

  \bibitem{FK}
  P.W. Kasteleyn and E.M. Fortuin, J. Phys. Soc. Jpn. Suppl. 26 (1969) 11; Physica 57 (1972) 536.
  
   \bibitem{Delfino}  
  G. Delfino,  M. Picco, R. Santachiara, and J. Viti, J. Stat. Mech. (2013) P11011.

  \bibitem{Duplantier2}
  B. Duplantier, Phys. Rev. Lett. {\bf 84}, 1363 (2000).
    
   \bibitem{Adams}
    D.~A. Adams, L.~M. Sander, and R.~M. Ziff, J. Stat. Mech. (2010) P03004.

  \bibitem{Corberi16}
  F. Corberi, E. Lippiello, and M. Zannetti, Europhys. Lett. {\bf  116}, 10006 (2016).

\bibitem{Nattermann87}
  T. Nattermann, Europhys. Lett. {\bf 4}, 1241 (1987).

\bibitem{Fisher86}
  M. E. Fisher, J. Chem. Soc., Faraday Trans. {\bf 82}, 1569 (1986).

\bibitem{Kardar87}
  M. Kardar, J. Appl. Phys. {\bf 61}, 3601 (1987).

\bibitem{Halpin89}
  T. Halpin-Healy, Phys. Rev. Lett. {\bf 62}, 442 (1989); Phys. Rev. A {\bf 42}, 711 (1990).

  \bibitem{Huse85}
    D. A. Huse and C. L. Henley, Phys. Rev. Lett. {\bf 54}, 2708 (1985).

    \bibitem{Kardar85}
      M. Kardar, Phys. Rev. Lett. {\bf 55}, 2923 (1985).

    \bibitem{Huse85b}
      D. A. Huse, C. L. Henley, and D. S. Fisher, Phys. Rev. Lett.
      {\bf 55}, 2924 (1985).

    \bibitem{Corberi16b}
      F. Corberi, E. Lippiello, and M. Zannetti,
      J. Phys. A: Math. Theor. {\bf 49}, 185001 (2016). 

    \bibitem{Loureiro10}
      M. P. O. Loureiro, J. J. Arenzon, L. F. Cugliandolo, and A. Sicilia,
      Phys. Rev. E {\bf 81}, 021129 (2010).

    \bibitem{Iwai93}
      T. Iwai, and H. Hayakawa, J. Phys. Soc. Japon {\bf 62}, 1583 (1993).

    \bibitem{Selke98}
      W. Selke, L. N. Shchur, and O. A. Vasilyev, Physica A {\bf 259}, 388 (1998).
      
    \bibitem{Heuer92}
      H.-O. Heuer, Phys. Rev. B {\bf 45} (1992) 5691.

    \bibitem{Kim94}
      J.-K. Kim, and A. Patrascioiu, Phys. Rev. Lett. {\bf 72}
      (1994) 2785.

\bibitem{Stinchcombe83}
       R. Stinchcombe, in: Phase Transitions and Critical Phenomena, vol.7, C. Domb
and J. L. Lebowitz, eds., (Academic Press, New York, 1983).

\end{thebibliography}
\end{document}